\documentclass[showpacs,twocolumn,pre,aps]{revtex4}
\usepackage[T1]{fontenc}        
\usepackage{graphicx}
\begin{document}
  \title{Ising model with periodic pinning of mobile defects}
  \author{M. \surname{Holtschneider}}
  \affiliation{Institut f\"ur Theoretische Physik, Technische Hochschule,D--52056 Aachen, Germany}
  \author{W. \surname{Selke}}
  \affiliation{Institut f\"ur Theoretische Physik, Technische Hochschule,D--52056 Aachen, Germany}

\begin{abstract}
A two--dimensional Ising model with short--range interactions
and mobile defects
describing the formation and thermal destruction of defect
stripes is studied. In particular, the effect of a local pinning
of the defects at the sites of straight equidistant lines is
analysed using Monte Carlo simulations and the transfer matrix
method. The pinning leads to a long--range ordered magnetic phase at
low temperatures. The dependence of the phase transition temperature, at
which the defect stripes are destabilized, on the pinning strength
is determined. The transition seems to be of first order, with
and without pinning.\\
\end{abstract}


\maketitle

\section{Introduction}

Striped magnetic structures in high--temperature superconductors and
related materials
have attracted much interest for more than a decade, both theoretically
and experimentally \cite{zaan,schulz,scheid1,tran,hoff,review}. In
that context, motivated by recent experiments on
(Sr, Ca, La)$_{14}$Cu$_{24}$O$_{41}$ \cite{buch1,kling}, a
class of rather simple two--dimensional Ising models has been
introduced describing the formation and thermal destruction
of defect stripes \cite{selke}.

The model consists of spin-1/2 Ising variables, mimicking Cu$^{2+}$
ions, and non--magnetic defects, $S=$ 0, corresponding to holes. The
spins are arranged in chains with antiferromagnetic
interactions, $J_a < 0$, between neighboring spins in adjacent
chains. Along the chains, neighboring spins are coupled
ferromagnetically, $J > 0$, while next--nearest neighbor spins
separated by a defect interact antiferromagnetically, $J_0 < 0$. The
defects are allowed to move along the
chain through the crystal. The mobility of the defects is determined by the
changes in the magnetic energy encountered during their
motion (annealed Ising model).

In a 'minimal variant' of the model, the couplings in the
chains, $J$ and $|J_0|$, are assumed to be indefinitely strong. The
minimal model has been shown to describe the formation of defect
stripes, oriented perpendicular to the chains, whose coherency
gets destroyed at a phase transition. At the transition, one
observes a pairing effect for the defects in the chains, reflecting
an effectively attractive
interaction between defects mediated by the magnetic interaction
between the chains, $J_a$. The thermal behavior of
the full model, choosing experimentally realistic values 
of the couplings in the chains, resembles closely that of the
minimal model \cite{selke}. 

The aim of this paper is to study the impact of a local defect
pinning energy of strength $E_p$ on thermal properties of the
minimal model. In the experimentally
studied (Sr, Ca, La)$_{14}$Cu$_{24}$O$_{41}$
compounds \cite{buch1,kling}, holes
are pinned by Ca-- or La--ions, which, in turn, are
rather immobile. In the following, we assume that the fixed pinning
sites form straight equidistant lines perpendicular to the chains, with 
the number of pinning sites being equal to the number of defects. Beyond
the specific experimental motivation, the model is hoped and
believed to be of genuine theoretical interest. 

Of course, the model still allows for thermal fluctuations of the 
defect stripes at finite pinning strength. Indeed, the instability of the
defect stripes and the effects of the
pinning on the spin ordering are intriguing features of the
present model. In particular, at low temperatures
spin correlations are expected to become long--ranged for
non--vanishing pinning, while they decay algebraically when
$E_p= 0$ \cite{selke}. The dependence of the phase transition, at
which the defect stripes get destroyed, on $E_p$
is an interesting aspect of the model as well. Without
pinning, the transition temperature had been estimated, but the type
of the transition had not been studied.

The layout of the paper is as follows. In the next section, we shall
introduce the model and the methods, Monte Carlo simulations
and transfer matrix calculations. Results
will then be presented and discussed in Sect. III. Finally, a
short summary concludes the article.\\    

\section{Model and methods}

\begin{figure}
\includegraphics[width=0.8\linewidth]{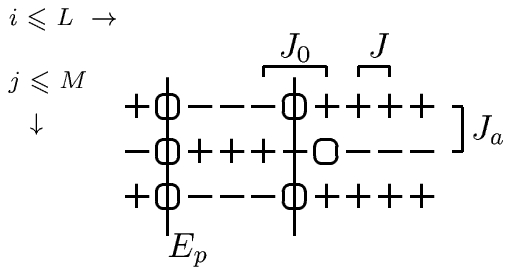}
\caption{Sketch of the interactions in the Ising model on
a square lattice with
periodic pinning of mobile defects.}
\label{fig1}
\end{figure}

\begin{figure*}
\includegraphics[width=\linewidth]{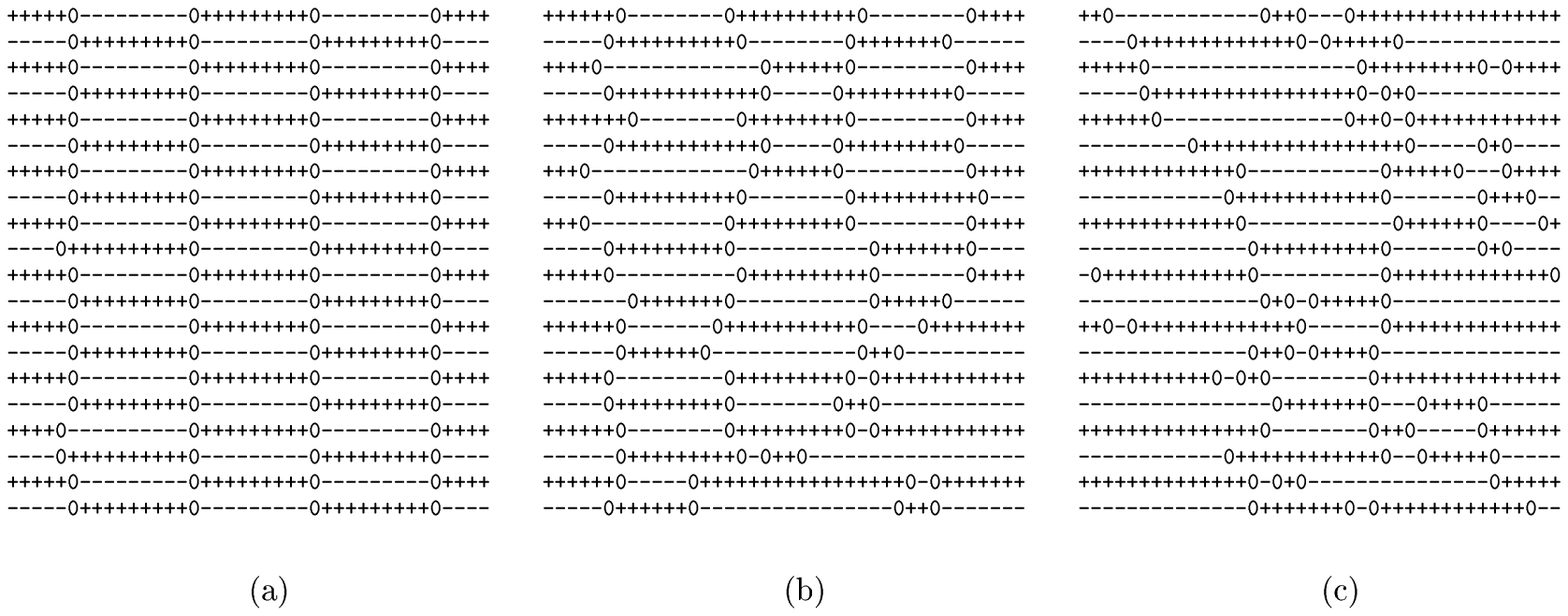}
\caption{Typical Monte Carlo equilibrium configurations
of the minimal model, $\Theta= 0.1$ and $q_p=$ 1.0, of
size $L= M=$ 40 at temperatures
$k_BT/|J_a|=$ 0.8 (a), 2.3 (b), and 2.9 (c). Only parts of
the systems are shown.}
\label{fig2}
\end{figure*}

We consider an Ising model on a square lattice, setting the lattice constant
equal to one. Each lattice site $(i,j)$ is occupied either by a
spin, $S_{i,j}=\pm 1$, or by a defect corresponding to
spin zero, $S_{i,j}= 0$, see Fig. 1. The defects
are mobile along one of the axes of the lattice, the
chain direction. The sites in the $j$-th chain are denoted
by $(i,j)$. We assume a ferromagnetic
coupling, $J > 0$, between neighboring
spins, $S_{i,j}$ and $S_{i \pm 1,j}$, along the chain, augmented by an
antiferromagnetic interaction, $J_{0} < 0$, between those next--nearest spins
in the same chain, which are separated by a defect. Spins in adjacent chains,
$S_{i,j}$ and $S_{i, j \pm 1}$, are coupled antiferromagnetically, $J_a < 0$.
Usually a minimal distance of two lattice spacings between
neighboring defects in a chain is assumed, i.e. two
defects are separated by at least
one spin due
to strong short range repulsion between defects (alternatively, one
may introduce an additional ferromagnetic coupling between 
spins separated by a pair of nearest--neighboring defects). A local 
pinning potential acts on the defects, lowering
the energy of the defects at fixed sites by
an amount $E_p$. In the following, we choose
pinning sites along equidistant straight lines, $i= i_p$, perpendicular
to the chains with the number of
pinning sites being equal to the number of defects, $N_d$.
Accordingly, the Hamiltonian of the model may be written as

\begin{eqnarray}
{\cal H} = -\sum\limits_{ij} [ J S_{i,j} S_{i \pm 1,j}
  + J_0 S_{i,j} S_{i \pm 2,j} (1- S_{i \pm 1,j}^2) \nonumber \\
  + J_a S_{i,j} S_{i,j \pm 1}
  + E_p (1- S_{i,j}^2)\delta_{i,i_p} ] ,
\end{eqnarray}
see Fig. 1. We assume that the
number of defects is the same in each chain, determined by the defect
concentration $\Theta$, denoting the total number of defects divided by the
total number of sites, $N_d/N$. In this study, we set $\Theta =0.1$, where
the distance between the pinning lines is then ten lattice spacings.

In the following we
consider the 'minimal' variant of the model by assuming
the couplings in the chain, $J$ and $|J_0|$, to be indefinitely
strong \cite{selke}. Thence the spins form intact clusters in the
chains between two consecutive defects, and neighboring
spin clusters have opposite sign. Thermal quantities depend only on, say,
$k_B T/|J_a|$ and the ratio $q_p= E_p/|J_a|$. 

To study the minimal model with pinning of
mobile defects, we used Monte Carlo techniques \cite{binder}
and the transfer matrix method \cite{transfer}.

In the simulations, a new configuration of spins and defects may
be generated by exchanging a defect with a neighboring
spin in a chain, reversing the sign of the spin to
keep intact spin clusters. The energy change associated with this elementary
process is determined by $J_a$ and $E_p$, see
the Hamiltonian (1). As
usual, the related Boltzmann factor determines the probability of
accepting the new configuration \cite{binder}.  Of course, simulations
are performed on
finite lattices with $N= L \times M$ sites, $L$ being the number of
sites in a chain. We shall present results for $L= M$. We employ
full periodic boundary conditions. To investigate
finite size effects, the linear dimensions, $L$ and $M$, were varied from 20
to 320. Typically, runs of at least
a few $10^6
$ Monte Carlo steps per defect were
performed, averaging then over such realizations to estimate
error bars. The pinning strength, $q_p= E_p/|J_a|$, ranged from 0 to 2.0.

The transfer matrix calculations were done in the standard way
\cite{transfer} with the matrices representing the interactions of 
the entire chains. All eigenvalues and eigenvectors were computed 
numerically, enabling us to derive quantities for arbitrary $M$,
being finite or infinite. Studying the case $\Theta= 0.1$, $L$ was 
chosen to be 20, with two defects per chain. Larger systems, 
i.e. with $L$ being at least 40, are outside the current reach 
of computer facilities. Of course, one may study the
case of more than two defect stripes in the case of $L=20$
by enlargening the defect concentration. We shall
consider here, however, only the case of a fixed value
of $\Theta= 0.1$. $q_p$ ranged from 0 to 5.0.   

Physical quantities of interest include the specific heat, $C$, and
spin correlation functions (depending, in general, on
the distance from the pinning lines, i.e. on $i$), parallel to the chains,

\begin{eqnarray}
 G_1(i,r)= \left(\sum\limits_{j} \langle S_{i,j} S_{i+r,j} \rangle \right)/L,
\end{eqnarray}
and perpendicular to the chains,

\begin{eqnarray}
 G_2(i,r)= \left(\sum\limits_{j} \langle S_{i,j} S_{i,j+r} \rangle \right)/L,
\end{eqnarray}
considering systems with $M= L$. Without pinning, the defects are
expected to be
delocalized so that there is full translational invariance, and the
spin correlations do not depend on $i$. Note 
that in the thermodynamic limit for infinitely large
distance, $r \rightarrow \infty$, the perpendicular correlations $G_2(r)$
determine the profile
of the squared magnetization

\begin{eqnarray}
 m^2(i)= \lim_{L \rightarrow \infty} m_L^2(i) = \lim_{L \rightarrow \infty} G_2(i,L/2)   
\end{eqnarray}

We also calculated
less common microscopic quantities which describe the stability of
the defect stripes and the ordering of the defects in the
chains. In particular, we computed the average
minimal distance, $d_m$, between each
defect in chain $j$, at position $(i_d,j)$, and those in the next
chain, at $(i_d',j+1)$, i.e.

\begin{eqnarray}
 d_m= \sum\limits_{i_d} \langle \min |i_d- i_d'| \rangle/N_d,
\end{eqnarray}
dividing the sum by the number of defects, $N_d$. Furthermore, we
calculated the cluster distribution, $n_d(l)$, denoting
the probability that consecutive defects in a chain are
separated by $l$ spins, in analogy
to the distribution of
cluster lengths in percolation theory \cite{stauffer}. Our main
emphasis will be on pairs of defects with $l= 1$. Finally, it
turned out to be quite useful to visualize the microscopic spin
and defect configurations as encountered during the simulation.

\section{Results}

\begin{figure}
\includegraphics[width=\linewidth]{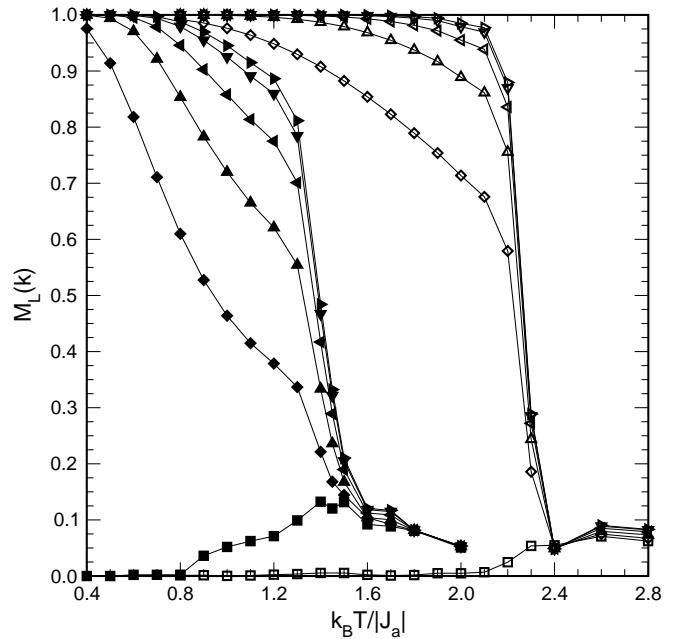}
\caption{Profiles of the absolute magnetization $M_L(k)$, at pinning strength
$q_p=$ 0.2 (full symbols) and 2.0 (open symbols),
with $k=$ 1 (squares), 2 (diamonds), 3 (triangles up), 4 (triangles left),
5 (triangles down) and 6 (triangles right). Results
have been obtained from simulations of systems of size
$L= M= 160$.}
\label{fig3}
\end{figure}

\begin{figure*}
\includegraphics[width=0.45\linewidth]{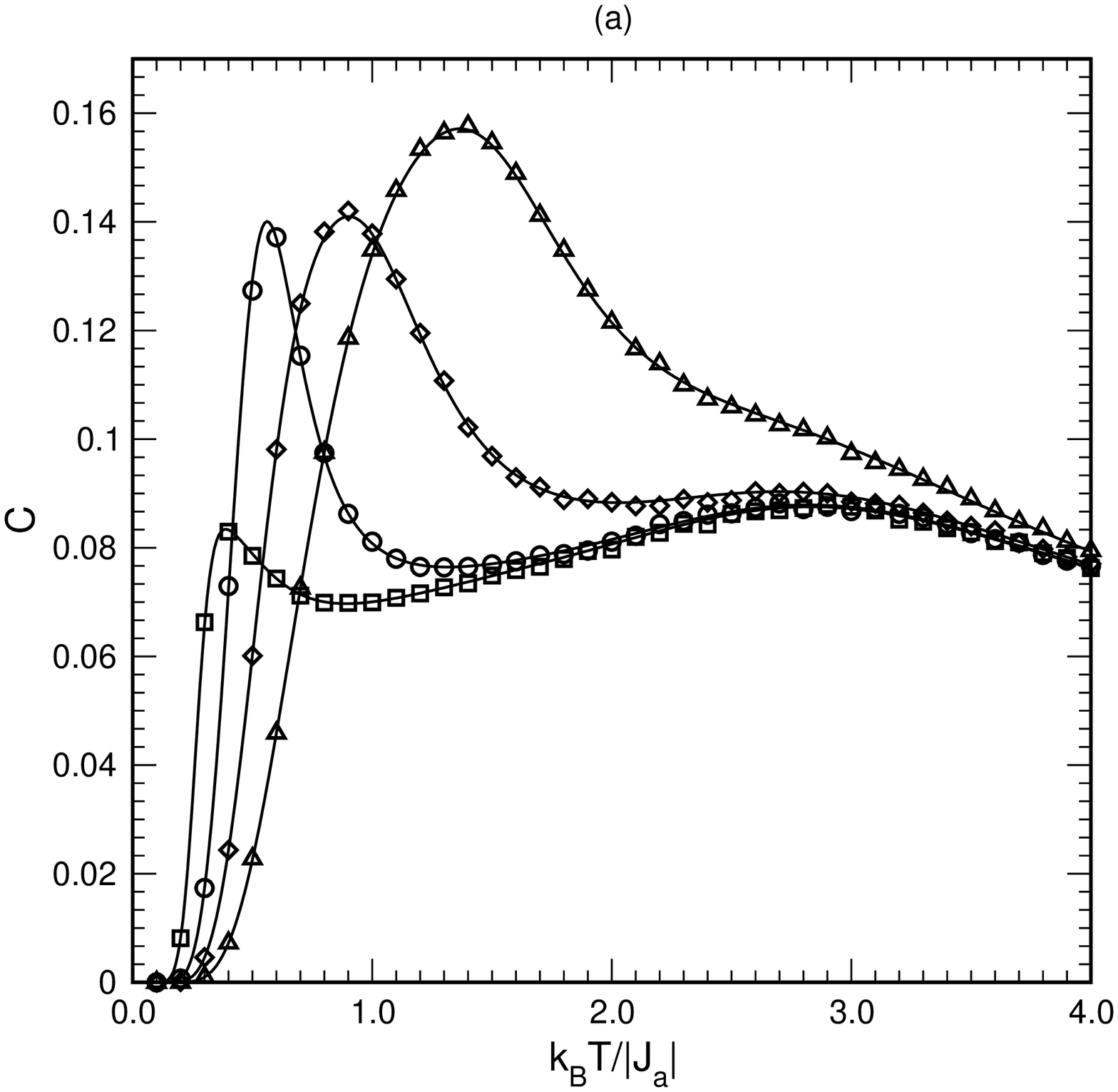}
\hspace{0.05\linewidth}
\includegraphics[width=0.45\linewidth]{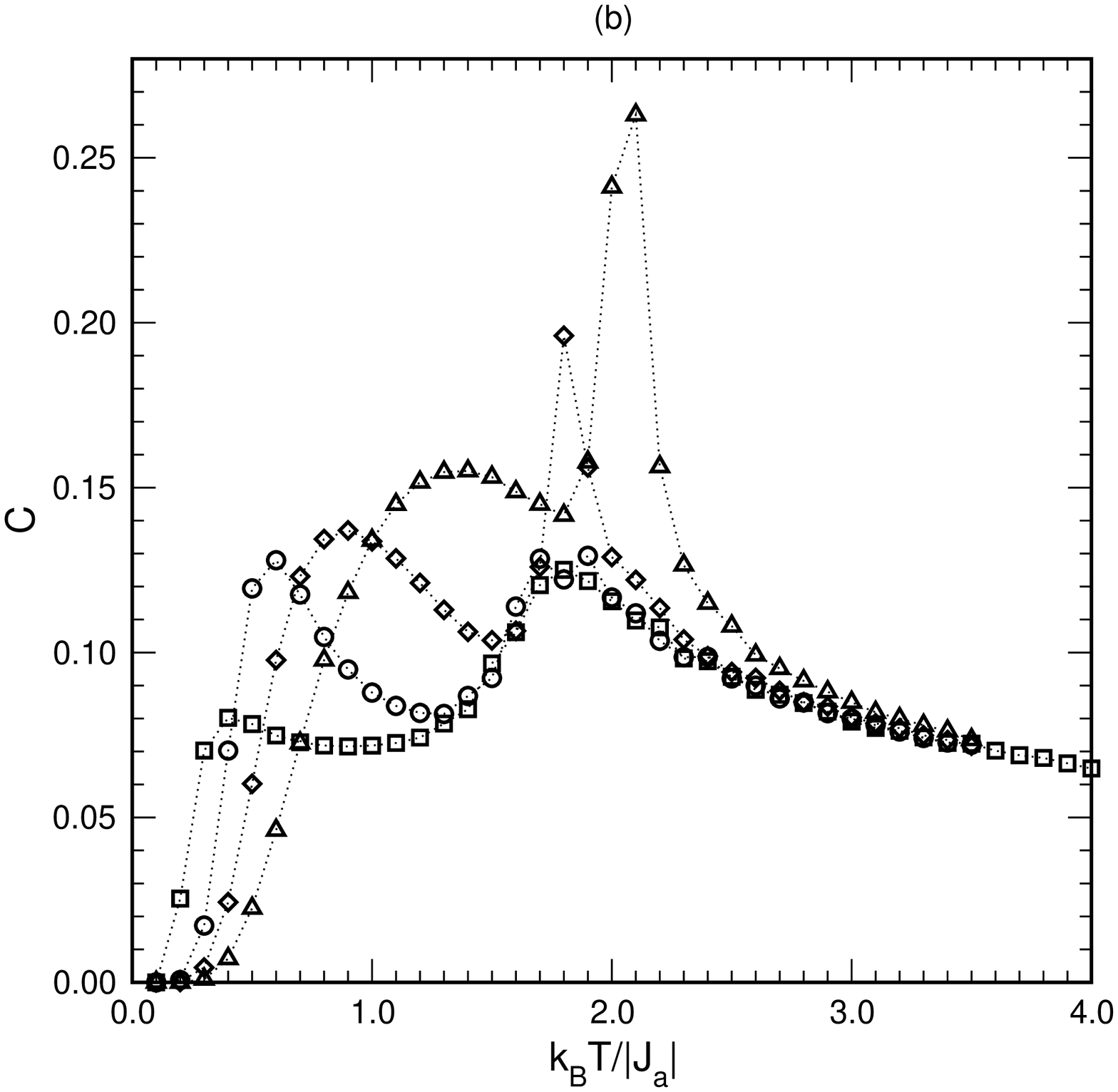}
\caption{Specific heat, $C$, at $q_p=$ 0 (squares), 0.2 (circles), 0.5
(diamonds), and 1.0 (triangles), for systems of size (a) $L= M=$ 20, showing
results from transfer matrix calculations (solid line) and simulations, and
of size (b) $L= M= 80$, obtained from simulations.}
\label{fig4}
\end{figure*}

In the ground state, $T= 0$, of the minimal
model, the defects form straight stripes
perpendicular to the chains, separating antiferromagnetic
domains of spins. Without pinning, $E_p= 0$, the ground
state is highly degenerate. Each arrangement of defect
stripes separated by at least two lattice spacings
has the same lowest possible energy, resulting in an exponential decay of the
correlations $G_1$ parallel to the chains, while the spins are perfectly
correlated perpendicular to the chains \cite{selke}. By introducing
the pinning potential, $E_p > 0$, the defect stripes coincide with
the pinning lines, at $i= i_p$. Obviously, $G_1$ continues to
oscillate, but now with a constant amplitude. Of course, the spin
correlations perpendicular
to the chains, $G_2(r)$, are equal to 1 for even distances $r$ and
$-1$ for odd distances $r$, when staying away from the
pinning lines, $i_p$.

Increasing the temperature, $T > 0$, the defects are
allowed to move so that the stripes start to meander
and finally break up, as exemplified in typical
Monte Carlo configurations depicted in Fig. 2. Due to the pinning
the defects tend to stick to the pinning lines at low
temperatures. The detachment or depinning of the defects from 
those lines is expected to occur without phase 
transition, as had been shown in the framework
of SOS models with pinning \cite{lipo}. The mapping of the minimal model
onto the standard SOS model has been discussed
before \cite{selke}. However, once the defects take positions
far from the pinning sites, the magnetic interactions
may mediate effectively attractive couplings between
the defects. As for vanishing pinning \cite{selke}, these
couplings, absent in standard SOS models, may eventually destroy the
coherency of the defect stripes
through a phase transition, as will be
discussed below. We shall provide numerical
evidence that the transition is of first order. The effect of the pinning
on the meandering and breaking up of the stripes, for various physical
quantities, is exemplified
in Figs. 3 to 7. Note that in most
of the figures we did not include error bars since they
were, typically, not larger
than the size of the symbols. Such a statement would not
hold for appreciably shorter Monte Carlo runs because of the rather slow fluctuations
of the defect stripes.

At $T > 0$, without pinning, $E_p= 0$, the model shows no
magnetic long--range order. The spin correlation function parallel
to the chains, $G_1$, has been shown, doing a free--fermion
calculation, to decay algebraically at
low temperatures \cite{selke}. Indeed, our new Monte Carlo
results both for $G_1$ and $G_2$ are consistent with
such an algebraic decay in
the low--temperature phase characterized by meandering defect
stripes whose positions can fluctuate rather freely. In particular, for
finite systems of size $L \times L$, the profile of the absolute
value of the magnetization, $|m_L(i)|= \sqrt{m_L^2(i)}$, reflects the
translational invariance, i.e. it does not depend
on $i$, and it decreases significantly with increasing system
size $L$. In marked contrast, with pinning, $E_p > 0$, at
low temperatures long--range magnetic order sets in, as seen
easily from the profiles of the absolute magnetization
between two pinning lines. The profiles are denoted in the
following by $M_L(k)$ with $k$ running from 1 to 11; $k= 1$ and
$k= 11$ denote the two pinning lines, the 
center line in between them is at $k= 6$. Obviously, one
has $M_L(12-k)= M_L(k)$ for reasons of symmetry. Examples 
of pertinent profiles are
displayed in Fig. 3 at weak, $q_p= 0.2$, and
strong, $q_p= 2.0$, pinning. Long--range order at low
temperatures follows from
the fact that the magnetization especially near the center between the
two pinning lines is largely independent of system
size. At high temperatures, the
magnetization decreases appreciably with
increasing system size, tending to zero in the
thermodynamic limit. Indeed, finite--size analyses allow one to 
locate the phase transition temperature as a function
of the pinning strength, $T_c(q_p= E_p/|J_a|)$. Estimates agree with
those obtained from analyses of the specific heat $C$, to
be discussed next. Note that $M_L(k)$ (or an average over these
absolute line magnetizations) may
be considered as the order parameter of the problem.

Results for the specific heat $C$ are depicted in Figs. 4a and 4b
for lattices with linear dimension $L=$ 20 and 80 at pinning
$0 \le q_p \le 1.0$. At fixed
pinning and varying temperature, one observes two
maxima in $C$. The
maximum at the lower temperature is almost independent of the
system size, and it stems from the meandering of the
defects stripes with few excitations, i.e. a small kink density, as
we checked by analysing and simulating corresponding SOS or  
TSK (terrace--step--kink) models \cite{lipo,selke2,einstein} with
pinning, similarly to the case without
pinning \cite{selke}. The lower maximum is
shifted towards higher temperatures when increasing the
pinning strength $E_p$. It may eventually be masked by the upper
maximum. The upper maximum of $C$, occurring at $T_{\text{max}} (L)$,
signals the instability of the defect stripes due to thermally
excited large fluctuations of the defect positions. At strong
pinning, these fluctuations are expected to set in once the defects
start to detach in significant numbers from the pinning lines, giving
then rise to a large specific heat, see Fig. 4b. In
any event, the height of the second maximum increases clearly
with increasing system size, indicating a phase transition
in the thermodynamic limit, $L \rightarrow \infty$. To
estimate the transition temperature, we plotted $T_{\text{max}}(L)$ versus
$1/L$, with $L$ going up to 160, see Fig. 5. From a linear extrapolation
one may approximate the phase transition
temperature $T_c(q_p)= T_{\text{max}} (L = \infty)$. $T_c(q_p)$ is found
to increase monotonically with $q_p$. More specifically, we obtain
the following estimates from the data
depicted in Fig.5.: $k_B T_c(q_p)/ |J_a| = 1.1 \pm 0.1$ at $q_p= 0.2$ (being close
to the estimate at $q_p= 0$ \cite{selke}), $1.30 \pm 0.1$ at $q_p=0.5$,
$1.55 \pm 0.1$ at $q_p= 1.0$, and $2.10 \pm 0.05$ at
$q_p= 2.0$, with  error bars reflecting some of the uncertainty in
the linear extrapolation. Finite size analyses
for other quantities lead to similar estimates
for the possible transition temperature, as already mentioned in
context of the magnetization profiles.

With pinning, the magnetization changes more and more drastically for larger
systems close to $T_{\text{max}}(L)$, compare to Fig. 3. This behavior may suggest
that in the thermodynamic limit the
phase transition is of first order, with a jump in the
magnetization at $T_c$. To clarify this aspect, we
determined the perpendicular correlation length, following from
$G_2$, when approaching $T_c$ from high temperatures. The correlation
length may be
estimated from analyzing the function \cite{upton}

\begin{eqnarray}
 \xi_{\text{eff}}(r)= -\left( \frac{d(\ln G_2(r))}{d(r)} \right)^{-1} \nonumber \\[5pt] \text{with} \quad G_2(r) = \sum\limits_{i} |G_2(i,r)| / L
\end{eqnarray}

Typically, the 'effective correlation length' $\xi_{\text{eff}}(r)$ increases
rather quickly monotonically for small $r$ until it
acquires a plateau--like behavior, and
finally it rises steeply due to the finite size effect and periodic
boundary conditions. Obviously, at a plateau of height $\xi_0$, one has
$G_2 \propto \exp (-r/\xi_0)$. Indeed, in the thermodynamic limit for
$T > T_c$, the height of the plateau at large $r$ obviously corresponds
to the standard correlation length $\xi$. Much care is needed close to
the transition because very large system sizes may have to be
studied to get an extended plateau. From simulations
of systems with $L= M= 160$, we determined the correlation
length versus temperature, at various fixed $q_p$. Using
linear extrapolation near $T_c(q_p)$, see above, we
estimate the perpendicular correlation length
at the transition. It is found to increase from about 20 lattice
spacings at $E_p/|J_a|= 2.0$ to about 30 lattice spacings
at $E_p= 0$; i.e., it is finite. This finding supports
the suggestion that the destruction of the defect stripes
occurs through a phase transition of first order, with and
also without pinning. A remark of caution may be added for
the case of vanishing pinning. There, spin
correlations in the low--temperature phase decay algebraically, and
one might expect a transition of Kosterlitz--Thouless type. As
has been noted before, however, algebraic order can be also destroyed
by a transition of first order \cite{halp,minn}.

The destruction of the defect stripes can be seen rather
directly in the average
minimal distance between defects in adjacent chains, $d_m$. In
Fig. 6, simulational data for system sizes $L=
M$ ranging from 20 to 160, at $q_p= 2.0$, are displayed. The
temperature dependence of $d_m$ resembles closely the one found
for the model without pinning \cite{selke}. While at
low temperatures $d_m(T)$
does not depend significantly on the system size, it starts to rise
rapidly at some characteristic temperature, corresponding
to $T_{\text{max}}(L)$ in the case of the specific heat, with
the height of the maximum in
the temperature derivative of $d_m$ increasing strongly with larger system
size. The location of the maximum, signalling the
breaking up of the stripes, moves to lower temperatures as $L$ gets
larger. The quantitative behaviour is quite similar to the one of
the specific heat and the magnetization
profiles, for the various pinning
strengths $q_p$ = 0, 0.2, 0.5, 1.0, and 2.0.

\begin{figure}
\includegraphics[width=\linewidth]{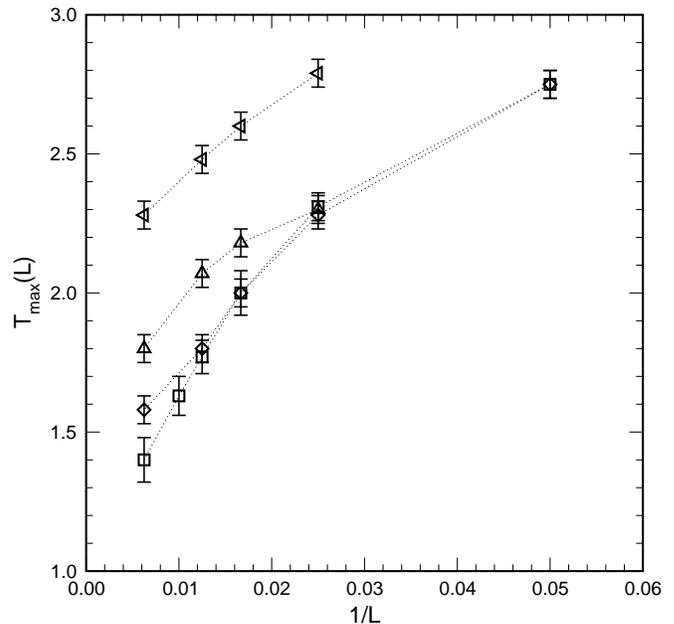}
\caption{Size dependence of the location of the maximum in the
specific heat, $T_{\text{max}}(L)$, as obtained from
simulations, at $q_p=$ 0.2 (squares), 0.5 (diamonds), 1.0
(triangles up), and 2.0 (triangles left) for $L= M$ ranging from 20 to 160.}
\label{fig5}
\end{figure}

The destabilization of the stripes seems to be driven
by effectively attractive couplings between consecutive
defects in a chain, mediated by the spin
interactions $J_a$ (possibly reminiscent of the spin--bag
mechanism \cite{schrieffer}). Indeed, effectively attractive couplings may
occur when two such defects, say, in chain $j$, at
sites $(i,j)$ and $(i+m,j)$, are displaced
strongly with respect to
corresponding defects in adjacent chains, $j \pm 1$, so that the spins in
those chains at sites in between $(i,j \pm 1)$ and $(i+m,j \pm 1)$
have the same sign as the spins between the two defects in
chain $j$. Such a situation may be realized, for
instance, when three defects
in chain $j$ are in a cage of four
defects in total, at, say, sites $(i, j \pm 1)$ and $(i +k, j \pm 1)$, in
the neighboring chains with spin clusters of
the same sign between the two pairs of defects in these chains 
$j-1$ and $j+1$. Then two of the three defects in the cage will move
towards each other \cite{selke}. In any event, due to the effectively
attractive coupling, mediated by $J_a$, two  
consecutive defects in chain $j$ tend to form a pair of next--nearest
neighboring defects having the minimal separation distance
of two lattice spacings. The temperature and
size dependence of the probability to find such
pairs of defects, given by the pair probability $n_d(l=1)$, is 
depicted in Fig. 7, at fixed pinning strength, $q_p= 2.0$, and
various system sizes. In general, the pronounced increase of the pair
probability occurs close to the temperature $T_{\text{max}}(L)$, where other
quantities signal the thermal instability
of the defect stripes as well. For larger system sizes the increase
in $n_d(1)$ gets sharper and sharper in
accordance with a transition
of first order. At strong pinning, the pair probability rises quite
drastically already in
systems of moderate size, see Fig. 7, possibly reflecting the moderate
correlation length at the transition, as discussed above.

Note that the type of stripe instability we observe here is not
included in standard descriptions of wall
instabilities in two dimensions \cite{halp,pokro,villain,pers}, where
either the number of walls is not fixed, giving rise to incommensurate
structures, or dislocations play an important role in the context of
melting of crystals. Also the bunching of steps in TSK models with
attractive step--step
interactions \cite{weeks} or instabilities in polymer filaments due to
attractive couplings \cite{burk,lipo2} are rather different from the
destruction of defect stripes due to the pairing of defects
induced by the inter--chain magnetic interactions $J_a$.

\section{Summary}

\begin{figure}
\includegraphics[width=\linewidth]{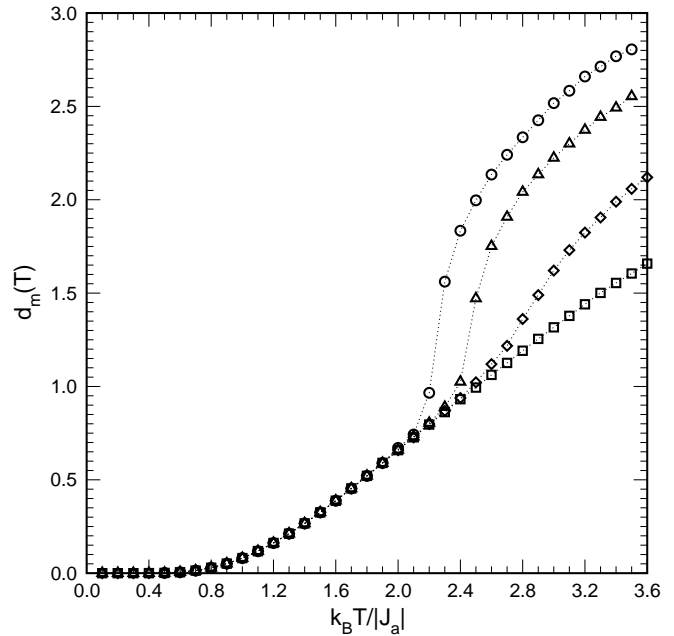}
\caption{Average minimal distance between defects in
adjacent rows $d_m(T)$, at $q_p=$ 2.0,
simulating systems of
size $L=M= 20$ (squares), 40 (diamonds), 80 (triangles), and
160 (circles).}
\label{fig6}
\end{figure}

\begin{figure}
\includegraphics[width=\linewidth]{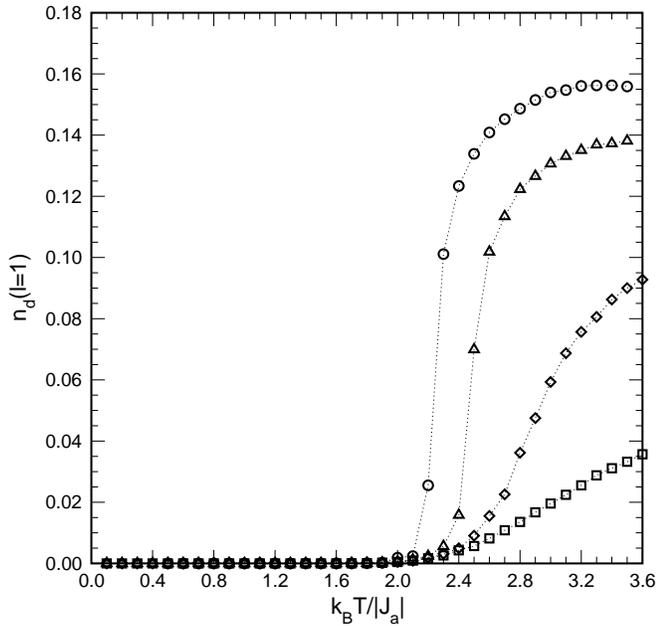}
\caption{Temperature dependence of the probability for
pairs of neighboring defects, $n_d(l=1)$, at
$q_p=$ 2.0, simulating systems
of size $L= M=$ 20 (squares), 40 (diamonds), 80 (triangles), and
160 (circles).}
\label{fig7}
\end{figure}

In this paper a two--dimensional Ising model with periodic local
pinning of mobile defects has been studied. Albeit the model
has been motivated by
recent experiments on cuprates with low--dimensional
magnetic interactions, the
model is believed to be of genuine theoretical interest as well.

In particular, based on  Monte Carlo
simulations and transfer matrix calculations, the model is found
to describe the pinning, meandering and, finally at higher
temperatures, the destruction of defect stripes.

The pinning gives rise to a 
long--range ordered magnetic phase at low temperatures while
magnetic correlations decay algebraically at
low temperatures without pinning.  

The thermal instability of the defect stripes, which
had been already identified for vanishing pinning, shifts towards
higher temperatures as the pinning strength increases. The instability
is signalled by pronounced anomalies, among others, in
the specific heat, in the magnetization
profile, in the probability of defect pairs with shortest
separation distance, and
in the average minimal distance between
defects in neighboring chains. The breaking up of the stripes
is caused by an effectively attractive coupling between
the defects mediated by the inter--chain interactions
between spins in adjacent chains. The attractive coupling leads to
a pairing of defects.

We provide evidence that the stripe instability 
results in a phase transition of first order, accompanied, in
the thermodynamic limit, by jumps in various quantities, including
the magnetization profile and the correlation length. This character
of the transition seems to persist for vanishing pinning.

\acknowledgments
It is a pleasure to thank B. B\"{u}chner, R. Klingeler, R. Leidl, V. L.
Pokrovsky, and S. Scheidl for very helpful suggestions and discussions.

\end{document}